\documentclass[twoside,twocolumn,9pt]{article}

\usepackage{dcolumn}
\usepackage{bm}
\usepackage{collcell}
\usepackage{siunitx}

\usepackage{extsizes}
\usepackage[super,sort&compress,comma]{natbib} 
\usepackage[version=3]{mhchem}
\usepackage[left=1.5cm, right=1.5cm, top=1.785cm, bottom=2.0cm]{geometry}
\usepackage{balance}
\usepackage{mathptmx}
\usepackage{sectsty}
\usepackage{graphicx} 
\usepackage{lastpage}
\usepackage[format=plain,justification=justified,singlelinecheck=false,font={stretch=1.125,small,sf},labelfont=bf,labelsep=space]{caption}
\usepackage{float}
\usepackage{fancyhdr}
\usepackage{fnpos}
\usepackage[english]{babel}
\addto{\captionsenglish}{%
  
}
\usepackage{array}
\usepackage{droidsans}
\usepackage{charter}
\usepackage[T1]{fontenc}
\usepackage[usenames,dvipsnames]{xcolor}
\usepackage{setspace}
\usepackage[compact]{titlesec}
\usepackage{hyperref}
\usepackage{amssymb}
\usepackage[normalem]{ulem}

\usepackage{epstopdf}

\definecolor{cream}{RGB}{222,217,201}

\begin{document}
\pagestyle{fancy}
\thispagestyle{plain}
\fancypagestyle{plain}{
\renewcommand{\headrulewidth}{0pt}
}

\makeFNbottom
\makeatletter
\renewcommand\LARGE{\@setfontsize\LARGE{15pt}{17}}
\renewcommand\Large{\@setfontsize\Large{12pt}{14}}
\renewcommand\large{\@setfontsize\large{10pt}{12}}
\renewcommand\footnotesize{\@setfontsize\footnotesize{7pt}{10}}
\makeatother

\renewcommand{\thefootnote}{\fnsymbol{footnote}}
\renewcommand\footnoterule{\vspace*{1pt}%
\color{cream}\hrule width 3.5in height 0.4pt \color{black}\vspace*{5pt}} 
\setcounter{secnumdepth}{5}

\makeatletter 
\renewcommand\@biblabel[1]{#1}            
\renewcommand\@makefntext[1]%
{\noindent\makebox[0pt][r]{\@thefnmark\,}#1}
\makeatother 
\renewcommand{\figurename}{\small{Fig.}~}
\sectionfont{\sffamily\Large}
\subsectionfont{\normalsize}
\subsubsectionfont{\bf}
\setstretch{1.125} 
\setlength{\skip\footins}{0.8cm}
\setlength{\footnotesep}{0.25cm}
\setlength{\jot}{10pt}
\titlespacing*{\section}{0pt}{4pt}{4pt}
\titlespacing*{\subsection}{0pt}{15pt}{1pt}

\fancyfoot{}
\fancyfoot[LO,RE]{\vspace{-7.1pt}\includegraphics[height=9pt]{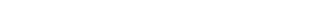}}
\fancyfoot[CO]{\vspace{-7.1pt}\hspace{11.9cm}\includegraphics{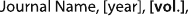}}
\fancyfoot[CE]{\vspace{-7.2pt}\hspace{-13.2cm}\includegraphics{head_foot/RF}}
\fancyfoot[RO]{\footnotesize{\sffamily{1--\pageref{LastPage} ~\textbar  \hspace{2pt}\thepage}}}
\fancyfoot[LE]{\footnotesize{\sffamily{\thepage~\textbar\hspace{4.65cm} 1--\pageref{LastPage}}}}
\fancyhead{}
\renewcommand{\headrulewidth}{0pt} 
\renewcommand{\footrulewidth}{0pt}
\setlength{\arrayrulewidth}{1pt}
\setlength{\columnsep}{6.5mm}
\setlength\bibsep{1pt}

\makeatletter 
\newlength{\figrulesep} 
\setlength{\figrulesep}{0.5\textfloatsep} 

\newcommand{\topfigrule}{\vspace*{-1pt}%
\noindent{\color{cream}\rule[-\figrulesep]{\columnwidth}{1.5pt}} }

\newcommand{\botfigrule}{\vspace*{-2pt}%
\noindent{\color{cream}\rule[\figrulesep]{\columnwidth}{1.5pt}} }

\newcommand{\dblfigrule}{\vspace*{-1pt}%
\noindent{\color{cream}\rule[-\figrulesep]{\textwidth}{1.5pt}} }

\makeatother

\twocolumn[
  \begin{@twocolumnfalse}

{\includegraphics[height=30pt]{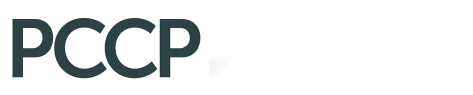}\hfill%
 \raisebox{0pt}[0pt][0pt]{\includegraphics[height=55pt]{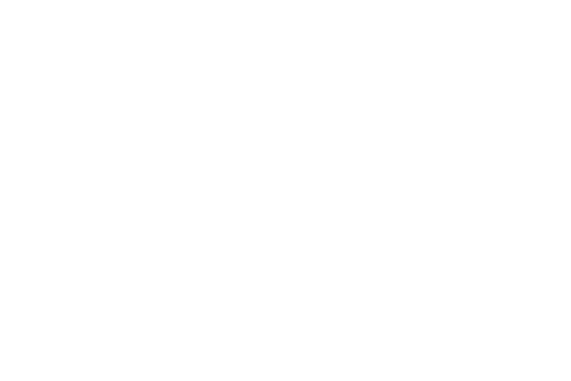}}%
 \\[1ex]%
 \includegraphics[width=18.5cm]{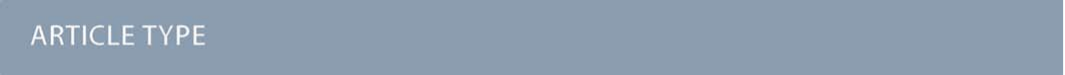}}\par
\vspace{1em}
\sffamily
\begin{tabular}{m{4.5cm} p{13.5cm} }

\includegraphics{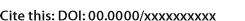} & \noindent\LARGE{\textbf{Relaxation dynamics in excited helium nanodroplets probed with high resolution, time-resolved photoelectron spectroscopy 
}} \\
\vspace{0.3cm} & \vspace{0.3cm} \\

 & \noindent\large{A. C. LaForge,$^{\ast,a}$ J. D. Asmussen,$^{\ast,b}$ B. Bastian,$^{b}$ M. Bonanomi,$^{c}$ C. Callegari,$^{d}$ S. De,$^{e}$ M. Di Fraia,$^{d}$ L. Gorman,$^{a}$ S. Hartweg,$^{f}$ S. R. Krishnan,$^{e}$ M. F. Kling,$^{g,h,i,j}$ D. Mishra,$^{a}$ S Mandal,$^{k}$ A. Ngai,$^{f}$ N. Pal,$^{d}$ O. Plekan,$^{d}$ K. C. Prince,$^{d}$ P. Rosenberger,$^{g,h}$ E. Aguirre Serrata,$^{a}$ F. Stienkemeier,$^{f}$ N. Berrah,$^{a}$ and M. Mudrich$^{b,e}$} \\
\vspace{5mm} 

\includegraphics{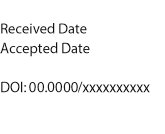} & \noindent\normalsize{Superfluid helium nanodroplets are often considered as transparent and chemically inert nanometer-sized cryo-matrices for high-resolution or time-resolved spectroscopy of embedded molecules and clusters. On the other hand, when the helium nanodroplets are resonantly excited with XUV radiation, a multitude of ultrafast processes are initiated, such as relaxation into metastable states, formation of nanoscopic bubbles or excimers, and autoionization channels generating low-energy free electrons. Here, we discuss the full spectrum of ultrafast relaxation processes observed when helium nanodroplets are electronically excited. In particular, we perform an in-depth study of the relaxation dynamics occurring in the lowest $1s2s$ and $1s2p$ droplet bands using high resolution, time-resolved photoelectron spectroscopy. The simplified excitation scheme and improved resolution allow us to identify the relaxation into metastable triplet and excimer states even when exciting below the droplets' autoionization threshold, unobserved in previous studies.
}

\\
\end{tabular}
 \end{@twocolumnfalse} \vspace{0.6cm}
  ]

\renewcommand*\rmdefault{bch}\normalfont\upshape
\rmfamily
\section*{}
\vspace{-1cm}

\footnotetext{$^{\ast}$~These authors contributed equally to this work.}
\footnotetext{$^{a}$~Department of Physics, University of Connecticut, Storrs 06269, Connecticut, 06269, USA.}
\footnotetext{$^{b}$~Department of Physics and Astronomy, Aarhus University, 8000 Aarhus C, Denmark, EU.}
\footnotetext{$^{c}$Istituto di Fotonica e Nanotecnologie, Consiglio Nazionale delle Ricerche, 20133 Milano, Italy, EU. }
\footnotetext{$^{d}$Elettra-Sincrotrone Trieste S.C.p.A., 34149 Basovizza TS, Italy, EU. }
\footnotetext{$^{e}$Department of Physics, Indian Institute of Technology-Madras, Chennai 600036, India.}
\footnotetext{$^{f}$Institute of Physics, University of Freiburg, 79104 Freiburg im Breisgau, Germany, EU.}
\footnotetext{$^{g}$Department of Physics, Ludwig-Maximilians-Universität Munich, 85748 Garching, Germany,EU.}
\footnotetext{$^{h}$Max Planck Institute of Quantum Optics, 85748 Garching, Germany, EU.}
\footnotetext{$^{i}$SLAC National Accelerator Laboratory, Menlo Park 94025, CA, USA.}
\footnotetext{$^{j}$Department of Applied Physics, Stanford University, Stanford 94305, CA, USA.}
\footnotetext{$^{k}$Indian Institute of Science Education and Research, Pune 411008, India.}

\section{Introduction}

Helium (He) nanodroplets are fascinating quantum fluid clusters with distinct properties compared to other types of atomic and molecular clusters. The constituent He atoms are loosely bound to one another by extremely weak attractive London dispersion forces and their light mass implies a large zero-point energy, i.e., the emergence of collective quantum behavior. Notably, He nanodroplets evaporatively cool to the ultralow temperature of 0.37~K, where they exhibit microscopic superfluidity~\cite{Grebenev:1998,Hartmann:1996,Tang:2002,Brauer:2013}. Although considered inert, upon excitation or ionization, He nanodroplets can become a highly reactive environment where numerous interatomic processes can occur~\cite{Mudrich2014}.

In general, electronic excitations in nanodroplets are mostly localized on single He atoms (He$^*$) which tend to form void cavities (`bubbles') due to the repulsive interaction between the excited electron and the surrounding He~\cite{Haeften:2001,Closser2010,Kornilov2011}. This leads to a broadening of the excited states in absorption spectra~\cite{Joppien1993}, although they largely retain their atomic character. We will refer to these broadened features as `bands' but we label them with their corresponding atomic electronic configuration. 
He$_2^*$ excited dimers may form, but the fraction of those directly formed by ultrafast association of an excited He$^*$ atom and a ground state He atom is very small~\cite{Ziemkiewicz2015,Closser2014}. The main formation mechanism of He$_2^*$ in large He droplets as well as in bulk superfluid He, tunneling into excited vibrational states of He$_2^*$, is much slower ($\sim 15~\mu$s) than any dynamics probed in the present experiment~\cite{keto1972dynamics,Buchenau:1991}. 

Quasi-free electrons in an excited nanodroplet occupy states in a `conduction band' $\gtrsim 1$~eV above the vacuum level~\cite{jortner1965study,Wang}. The corresponding electronic wavefunctions are localized in the interstitial spaces between He atoms~\cite{Borghesani:2007}. Autoionization of these highly-excited states then leads to the emission of small charged He clusters He$_n^+$, $n=2,\,3,\dots$ and low-energy electrons~\cite{BuenermannJCP:2012,Peterka2003}.


The precise electronic structure of excited He nanodroplets still remains to be fully resolved~\cite{Closser2010,Ziemkiewicz2015}. In particular, the dynamics of electron localization, relaxation, and atomic rearrangements induced by the electronic excitations are subjects of ongoing research~\cite{Closser2014}. The ultrafast dynamics of excited He nanodroplets has been studied in several femtosecond pump-probe experiments employing both laser-based high-harmonic generation (HHG) sources~\cite{Ziemkiewicz2015} and extreme-ultraviolet (XUV) free-electron lasers (FELs)~\cite{Mudrich,asmussen2021unravelling}. Essentially, ultrafast localization of the excitation on single atomic sites was confirmed, followed by the electronic relaxation into metastable states, and the formation of nanoscale bubbles around the excited atoms on the timescale of $\sim 1$~ps. In the first HHG-based experiments where near-infrared probe pulses were used, ejection of He$^*$ excited atoms was observed~\cite{Kornilov2011}. Subsequent experiments by the same group showed that UV pulses are better suited for probing the full He droplet dynamics and that the ejection of a He$^*$ atom actually is a minor relaxation channel~\cite{Ziemkiewicz2014,Ziemkiewicz2015}.
The lowest excited atomic state of He, $1s2s\,^3S$, has only been observed at photon energies $h\nu\gtrsim 23~$eV, where He droplets autoionize and optically inaccessible states can be populated through electron-ion recombination~\cite{VonHaeften1997,asmussen2021unravelling}. Static measurements have shown that even highly excited He droplet states efficiently relax into the lowest excited $1s2s\,^{1,\,3}S$ atomic states~\cite{BenLtaief2019}. Eventually, He$_2^*$ excimers form at the He droplet surface, detaching from the droplets over the course of vibrational relaxation~\cite{Buchenau:1991,VonHaeften1997}. 

In this article, we present an overview of the XUV-pump, UV-probe dynamics of resonantly excited He nanodroplets over a wide range of excitation energies and electron energies. Subsequently, we focus on the electronic relaxation processes occurring in the lowest $1s2p$ and $1s2s$ droplet bands~\cite{Joppien1993}. Our previous measurements~\cite{Mudrich,asmussen2021unravelling} have been limited by the spectrometer resolution and by experimental noise from XUV/UV stray light. Here, with improved resolution and a higher signal-to-noise ratio, we are able to observe new features due to relaxation in the time-resolved photoelectron spectra. In particular, when exciting the system to the lowest optically allowed droplet states, $1s2s\,^1S$ and $1s2p\,^1P$, we observe efficient relaxation into triplet atomic states, indicating droplet-induced spin-relaxation as well as population of the first excimer state of He$_2^*$. Compared to previous work at higher excitation energies~\cite{Ziemkiewicz2015,asmussen2021unravelling}, the relaxation process discussed here is simpler since it only involves intra-band relaxation as opposed to a combination of inter- and intra-band relaxations. Thus, with a simpler excitation scheme and improved resolution, we are able to better distinguish weak features in the photoelectron spectrum and to probe their ultrafast dynamics on femtosecond timescales.      

\section{Experimental methods}

The pump-probe experiment was performed at the Low Density Matter (LDM) endstation~\cite{Lyamayev2013} of the seeded FEL FERMI, in Trieste, Italy~\cite{Allaria2012,Allaria2012a}. The FEL pump pulse, operated at a repetition rate of 50~Hz, was primarily tuned to a photon energy $h\nu_1$ near the $1s2p$ droplet resonance ($21.6$~eV)~\cite{Joppien1993} via the seed laser and undulator gaps, yielding a pulse length $t_p\approx 100$~fs (full width at half maximum, FWHM)~\cite{finetti2017pulse}. The FEL pulse intensity was adjusted to limit the effects of multiphoton ionization~\cite{Ovcharenko:2014,LaForge:2014,Ovcharenko:2020} with typical values of $I_{XUV}\approx 1\times10^{10}$~W/cm$^{2}$, as derived by the pulse energy measured upstream by gas ionization, taking into account the nominal reflectivity of the optical elements in the beam transport system~\cite{Svetina2015}. The UV probe pulse was obtained from a frequency-tripled Ti:Sapphire laser ($h\nu_2\,=\,4.65$~eV) with a pulse intensity of $I_{UV}\approx 1\times10^{11}$~W/cm$^{2}$~\cite{Finetti2017}. A tin filter of 200\,nm thickness was used to suppress contributions from higher order harmonic radiation whenever the pump photon energy $h\nu_1$ was tuned to $21.0$~eV and $21.6$~eV. The cross correlation between the FEL and the probe laser was 200~fs FWHM, measured by resonant (1+1') two-photon ionization of atomic He,where 1' indicates one-photon absorption by the probe pulse~\cite{Finetti2017}. 

A supersonic jet of He nanodroplets was produced by expansion of high pressure He gas (50~bar) through a pulsed, cryogenically cooled (14~K) Even-Lavie-type nozzle (50~$\mu m$). From the expansion conditions (backing pressure and nozzle temperature), the mean droplet size was estimated to $N_\mathrm{He}\approx 10^5$ He atoms per droplet~\cite{Toennies2004a,Gomez:2011}. After expansion, the nanodroplet beam passed through a 1~mm skimmer and traversed approximately 90~cm to the interaction region. The nanodroplet beam was perpendicularly crossed by the FEL and UV beams at the center of a high resolution ($E/\Delta E\approx 50$) magnetic bottle electron spectrometer~\cite{Squibb2018}. 

\section{Results and Discussion}

\begin{figure*}[t]
    \centering
    \includegraphics[width=1\textwidth]{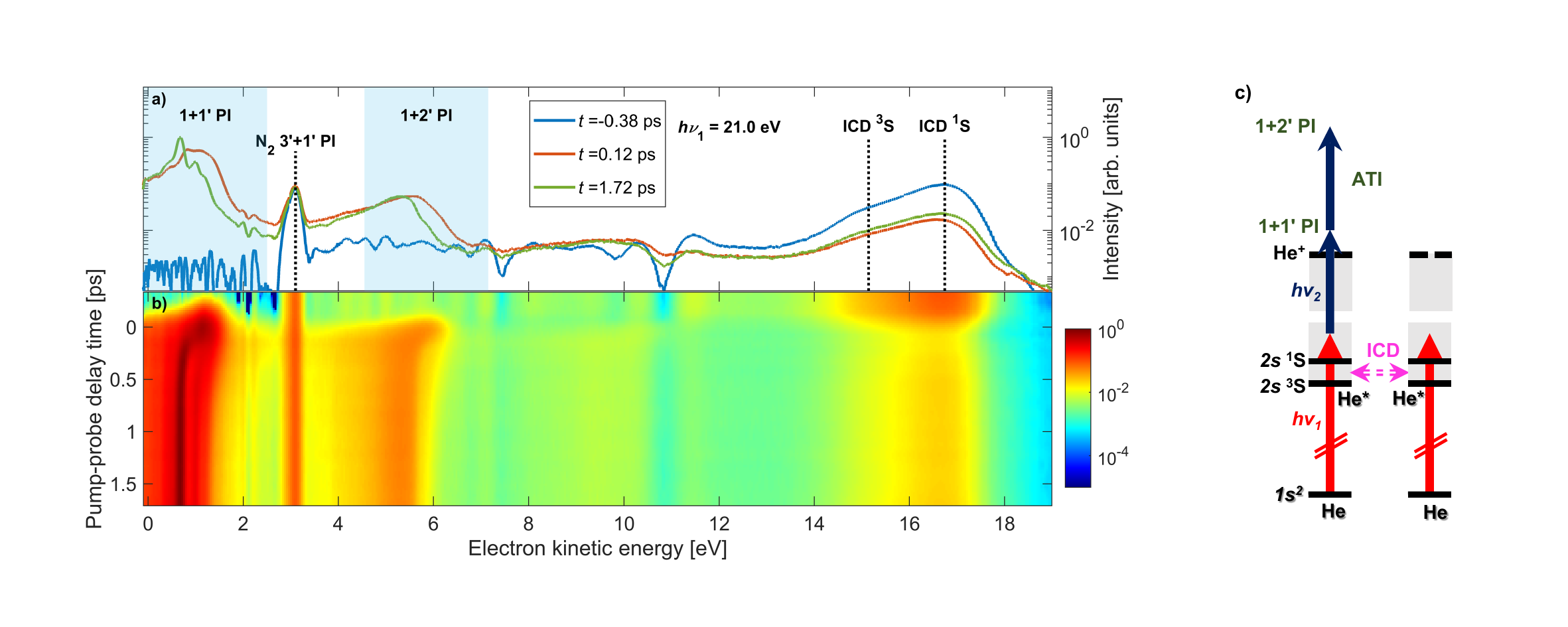}
    \caption{Overview of the time-resolved photoelectron spectra of He nanodroplets measured with XUV-pump and UV-probe pulses. The pump photon energy is $h\nu_1 = 21.0$~eV and the probe photon energy is $h\nu_2 =4.65$~eV. Panel a) shows photoelectron spectra at selected pump-probe delays on a logarithmic scale; Panel b) shows a map composed of 28 spectra at different pump–probe delays. The shaded areas in a) indicate the regions in the electron spectrum where 1-2 UV probe photons are absorbed by the XUV-excited He droplets. The vertical dashed lines mark the energies of electrons created by ICD in He nanodroplets assuming atomic excitation energies, and $3'+1'$-photoionization of N$_2$. Panel c) shows an energy level diagram illustrating transitions induced by the pump ($h\nu_1$) pulse (red vertical arrows) and the probe ($h\nu_2$) pulse (blue arrows). The ICD process, where one excited He$^*$ atom decays to the ground state and another He$^*$ is ionized, is depicted by the horizontal pink arrow. The ATI process, where an electron is promoted to high kinetic energies by absorption of one or more photons in the ionization continuum, is represented by the upper blue vertical arrow.
    \label{fig:full_spec_21d0eV}}
\end{figure*}

\begin{figure*}[t]
    \centering
    \includegraphics[width=0.9\textwidth]{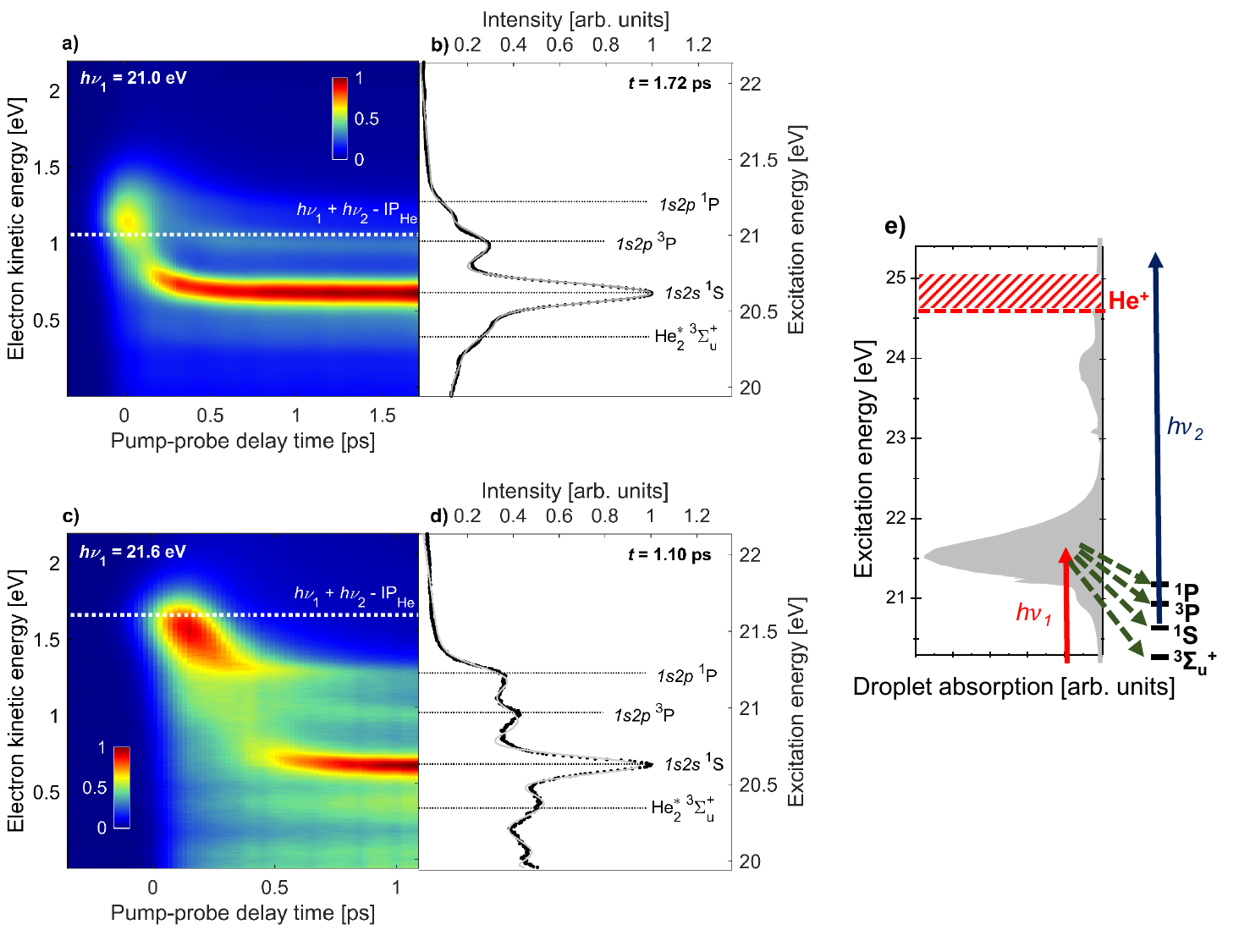}
    \caption{Time-resolved photoelectron spectra of He droplets measured with XUV-pump and UV-probe pulses at photon energies $h\nu_1 = 21.0$~eV [a) and b)] and $h\nu_1 = 21.6$~eV [c) and d)]. The maps [a) and c)] are composed of 28 and 31 spectra at different pump–probe delays, respectively. The white dashed lines indicate the expected electron energies for direct 1+1' photoionization. Panels b) and d) show the electron spectra at fixed delays of 1.7~ps and 1.1~ps, respectively. The black horizontal lines indicates the expected electron energies for photoionization of specific excited states of quasifree He$^*$ atoms and He$_2^*$ excimers. The gray line is the multi-Gaussian fit to the experimental spectra. See the main text for additional details. Panel e) shows the He droplet absorption spectrum from Joppien et al.~\cite{Joppien1993} and the energy levels populated in the course of relaxation. The red and blue arrows represent the pump and probe pulses, respectively, and the green arrows represent the relaxation channels.}
    \label{fig:21d6eV+21d0eV}
\end{figure*}

When a He nanodroplet is resonantly excited by an intense XUV pulse, a wide variety of dynamic processes is initiated, including internal relaxation, interatomic energy transfer, and multiphoton ionization. To probe the dynamics, we use a time-delayed UV laser pulse which directly ionizes the excited nanodroplets and we measure the energy of emitted electrons. In this way, we track the dynamics of the various processes as a function of the pump-probe delay. Figure~\ref{fig:full_spec_21d0eV}~a) shows representative raw electron kinetic energy distributions measured for He nanodroplets that are resonantly excited into the lowest optically allowed droplet state $1s2s\,^1S$ at $h\nu_1=21.0$~eV for selected pump-probe delays. Fig.~\ref{fig:full_spec_21d0eV}~b) shows a map of the electron kinetic energy distributions ($x$-axis) as a function of the pump-probe delay time ($y$-axis). To better understand the spectral features, Fig.~\ref{fig:full_spec_21d0eV}~c) shows an energy-level diagram of He atoms (horizontal black lines) and He droplets (gray shaded areas). The vertical arrows depict photo-excitation (red) and photo-ionization steps (black). The horizontal pink arrow indicates autoionization of pairs of excited He atoms in one droplet which is discussed in detail below.

Overall, numerous dynamic features can be observed in the spectrum in different ranges of the electron energy. At low electron energies $\lesssim 2~$eV, we observe prominent features shifting to even lower energies as the delay increases. These are generated by 1+1' photoionization through singly excited states of the droplets. The shifting of the peaks reflects the relaxation of the excited states into low-lying, metastable states, as previously studied for the $1s2p$ band~\cite{Mudrich} and the $1s3p$ band~\cite{asmussen2021unravelling}. Additionally, we observe a similar feature at slightly higher kinetic energies around 5~eV, which is due to the excited state absorbing two UV photons from the probe pulse resulting in 1+2' `above-threshold ionization' (ATI) in the excited system. 
ATI of multiply excited He nanodroplets by near-infrared or visible probe pulses has recently been found to be strongly enhanced compared to isolated excited He atoms, pointing at a collective coupling effect~\cite{michiels2021enhancement}. 
At electron energies around 3~eV, there is a time-independent peak, which we attribute to the photoionization of background nitrogen molecules by the probe pulse through a 3'+1' REMPI scheme. Note that the peak cancels out when the background gas spectrum is subtracted; Therefore, it is not related to the He nanodroplet beam.

The feature at electron energies around 16~eV is due to autoionization of pairs of excited He atoms in one droplet according to a process known as ICD~\cite{Cederbaum:1997,Kuleff:2010}. For this particular case, ICD occurs via the process $\mathrm{He}^* + \mathrm{He}^* \rightarrow \mathrm{He} + \mathrm{He}^+ + e_\mathrm{ICD}$ which manifests itself in the emission of electrons $e_\mathrm{ICD}$ with a characteristic energy. For the given experimental parameters, we estimate the number of resonantly excited He$^*$ atoms per He droplet to $N_*=N_\mathrm{He}\sigma_{XUV} t_p I_{XUV}/h\nu_1 \approx 700$, where $\sigma_{XUV}\approx 25~$Mbarn~\cite{Ovcharenko:2014,Ovcharenko:2020}. This leads to an estimated number of photoelectrons emitted out of the lowest-lying He excited states $N_e^{PI} = N_*\sigma_{UV} t_p I_{UV}/h\nu_2 \approx 100$, given that the photoionization cross sections of these states range from $\sigma_{UV}=4$ to $7~$Mbarn~\cite{chang1995effect}. The number of electrons created by ICD can be estimated based on the previously measured ICD efficiency, $p_{ICD}\approx 30$~\%, for a droplet excitation rate of $N_*/N_\mathrm{He} \approx 0.7~$\%~\cite{laforge2020time}, yielding $N_e^{ICD} = N_* \times p_{ICD}/2 \approx 100$. The factor $1/2$ accounts for the fact that it takes a pair of He$^*$ atoms to create an e$_ICD^-$. Thus, we expect to observe roughly an equal amount or photoelectrons and ICD electrons, in agreement with the experimental finding (Fig.~\ref{fig:full_spec_21d0eV}).

At positive pump-probe delays this ICD process is partly quenched because the probe pulse depletes the population of $\mathrm{He}^*$ excited states prior to their decay. From the rise of the ICD signal at delays $\gtrsim 0.2$~ps we previously inferred an ICD lifetime on the order of a few hundred fs~\cite{laforge2020time}. We observed only a weak dependence of the ICD lifetime on the degree of He droplet excitation which is controlled by the photon flux $I_{XUV}/h\nu_1$ and the droplet size $N_\mathrm{He}$. This finding was rationalized by a model in which the bubbles forming around $\mathrm{He}^*$ excitations can merge, thereby accelerating two nearby $\mathrm{He}^*$ atoms toward each other~\cite{laforge2020time}. Thus, the detected ICD lifetime is mainly determined by the quantum fluid dynamics of the merging bubbles, rather than by the distance-dependent ICD rate for a bare pair of $\mathrm{He}^*$ atoms.
When the He nanodroplet is strongly excited with more intense FEL pulses, it can spontaneously evolve into a nanoplasma state by `collective autoionization' (not shown in these spectra)~\cite{Ovcharenko:2014,LaForge:2014,Ovcharenko:2020}. In that case the electron spectrum is dominated by a distribution of low-energy electrons due to thermal emission.

\begin{figure}[t]
    \centering
    \includegraphics[width=0.95\columnwidth]{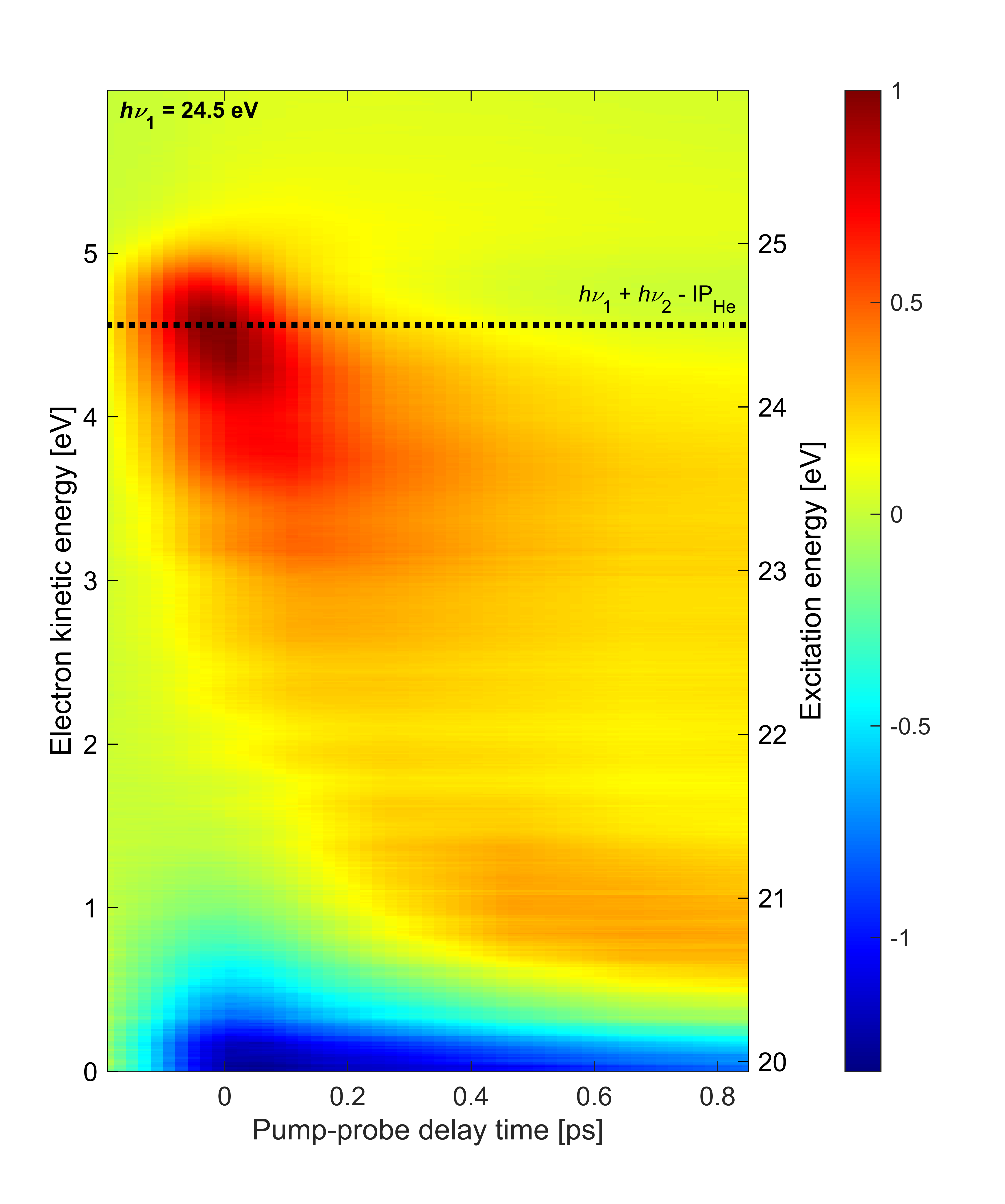}
    \caption{Time-resolved photoelectron spectra recorded at $h\nu_1 = 24.5$~eV composed of 13 spectra at different pump-probe delays. The dashed line indicates the expected electron energy for direct 1+1' photoionization.}
    \label{fig:24d6eV}
\end{figure}

For the remainder of this article, we will specifically focus on the relaxation dynamics of resonantly excited droplets, which are observed via 1+1' photoionization at low electron energies. These measurements were performed with a high-resolution magnetic bottle spectrometer enabling us to distinguish new features, which previously remained unobserved~\cite{Mudrich}. Fig.~\ref{fig:21d6eV+21d0eV} shows the electron spectra ($y$-axis) as a function of pump-probe delay ($x$-axis) for two photon energies, $h\nu_1 = 21.0$~eV and 21.6~eV in a) and c), respectively.
The zero of the pump-probe delay, $t_0$, is determined from the falling edge of the integrated ICD signal as it is quenched by the probe pulse.
Fig.~\ref{fig:21d6eV+21d0eV}e) shows the absorption spectrum of He droplets recorded by Joppien \textit{et al.}~\cite{Joppien1993} by means of fluorescence detection along with He$^*$ energy levels and arrows indicating the relaxation pathways following excitation. The two pump photon energies, $h\nu_1 = 21.0$~eV and $21.6$~eV, correspond to excitation into the low-energy shoulder on the main absorption band and to the maximum of this absorption band, respectively. The droplet excitation at $h\nu_1 = 21.0$~eV correlates with the $1s2s\,^{1}S$ atomic state~\cite{Joppien1993,Closser2010}, which is inaccessible by electric dipole transition in the atom. It is more likely to occur near the surface of the droplet where the inversion symmetry is broken~\cite{Mudrich}. Excitation into the main absorption peak ($h\nu_1 = 21.6$~eV) can mostly be associated with the $1s2p\,^{1}P$ atomic state~\cite{Joppien1993,Closser2010}. 
Additionally, Figs.~\ref{fig:21d6eV+21d0eV} b) and d) display electron spectra at longer delays when the fast relaxation dynamics have subsided. For these spectra, as well as the results in Fig.~\ref{fig:24d6eV}, we have subtracted the background to better observe the dynamic features. Specifically, we subtract, for each time delay, the electron spectrum produced by the FEL pulse alone, \textit{i.\,e.} in the absence of the UV pulse. 

We observe some similar features for the two photon energies. Near zero delay, the electron energy is at its highest and matches the combined photon energies of the XUV-pump and UV-probe pulses reduced by the ionization energy of He, $h\nu_1 + h\nu_2 - IP_\mathrm{He} = 1.1$~eV for $h\nu_1 = 21.0$~eV (white dashed line in Fig.~\ref{fig:21d6eV+21d0eV} a) and 1.7~eV for $h\nu_1 = 21.6$~eV (white dashed line in c). For positive delays we observe a fast relaxation to lower energies. For $h\nu_1 = 21.0$~eV, the relaxation is so fast that it appears as an almost vertical band, meaning relaxation proceeds within the experimental time resolution (200~fs). For $h\nu_1 = 21.6$~eV, the relaxation is slower resulting in a clearly visible shift from 1.7 to 1.3~eV over about 300~fs. This shift is attributed to intraband electronic relaxation towards the lower edge of the droplet band correlating to the $1s2p\,^1P$ state of He. Interestingly, we see a significant narrowing of this band at delays $\gtrsim 200$~fs converging toward a sharp peak whose position would match the $1s2p\,^1P$ He atomic state, see Fig.~\ref{fig:21d6eV+21d0eV}~d). Simultaneously, new sharp features appear at lower electron kinetic energies. The most prominent one is an intense, nearly horizontal band at electron kinetic energy $eKE\approx 0.6$~eV in Fig.~\ref{fig:21d6eV+21d0eV}~c). It was observed in our previous work~\cite{Mudrich} and was attributed to relaxation into the $1s2s\,^{1}S$ droplet state. This relaxation is likely mediated by a crossing of potential energy curves of the He$_2^*$ excimer~\cite{fiedler2014interaction}. It is accompanied by the formation of a void bubble around the localized He$^*$ excitation~\cite{VonHaeften2002}. At later times, this bubble migrates to the droplet surface and releases an excited He$^*$ atom or a He$_2^*$ excimer which either remains weakly bound to the droplet surface or fully detaches into the vacuum~\cite{Buchenau:1991}. These species have previously been detected by static fluorescence measurements~\cite{VonHaeften1997}. 

Besides the $1s2\,s^{1}S$ atomic state, additional features are clearly visible, which were not observed in our prior work~\cite{Mudrich}. 
To gain a better understanding of the origin of these features we offset the vertical right axes in Figs.~\ref{fig:21d6eV+21d0eV}~a) and c) to correspond to the excitation energy of the system. In this way, we can visually assign the two horizontal bands at higher kinetic energies to the $1s2p\,^3P$ and $1s2p\,^1P$ atomic states, respectively. Note that in a different experiment where we studied the relaxation dynamics in He nanodroplets excited to the $1s3p$ band~\cite{asmussen2021unravelling}, we did observe a shoulder on the $1s2s\,^1S$ atomic peak, which we now fully resolve and attribute the peak to the $1s2p\,^3P$ atomic state. For the static measurements, shown in Figs.~\ref{fig:21d6eV+21d0eV}~b) and d), 
horizontal lines are added to visualize the different excited states. These peaks are most prominent for $h\nu_1 = 21.6$~eV. For 21.0~eV, the lower intensity of the $1s2p\,^1P$ atomic state could be due to its nominal excitation energy being slightly higher than the FEL photon energy. 

We additionally observe another feature at lower kinetic energy, which we attribute to the $^{3}\Sigma^{+}_{u}$ state of the He$^*_2$ excimer photoionized into the lowest vibrational state $v=0$ of the He$_2^+$ dimer ion. For photoionization into the first excited vibrational state $v=1$ of He$_2^+$ we expect the electron energy to be $eKE\approx 0.17~$eV, which roughly matches the position of the asymmetric feature peaked around $eKE=0.1~$eV in the electron spectrum at $h\nu_1=21.6~$eV excitation. This feature may also have contributions from autoionization of superexcited He droplets~\cite{Peterka:2003}; Quasi-bound states $\lesssim 1.1$~eV above the ionization energy of atomic He, $E_i=24.59$~eV, or states bound by $\lesssim 1.6$~eV below $E_i$, populated by `re-excitation' out of the dark state $1s2s\,^3S$, can decay by emission of low-energy electrons~\cite{Kornilov2011,asmussen2021unravelling}. However, we refrain from a conclusive assignment of this feature. 

\begin{figure}[t]
    \centering
    \includegraphics[width=0.95\columnwidth]{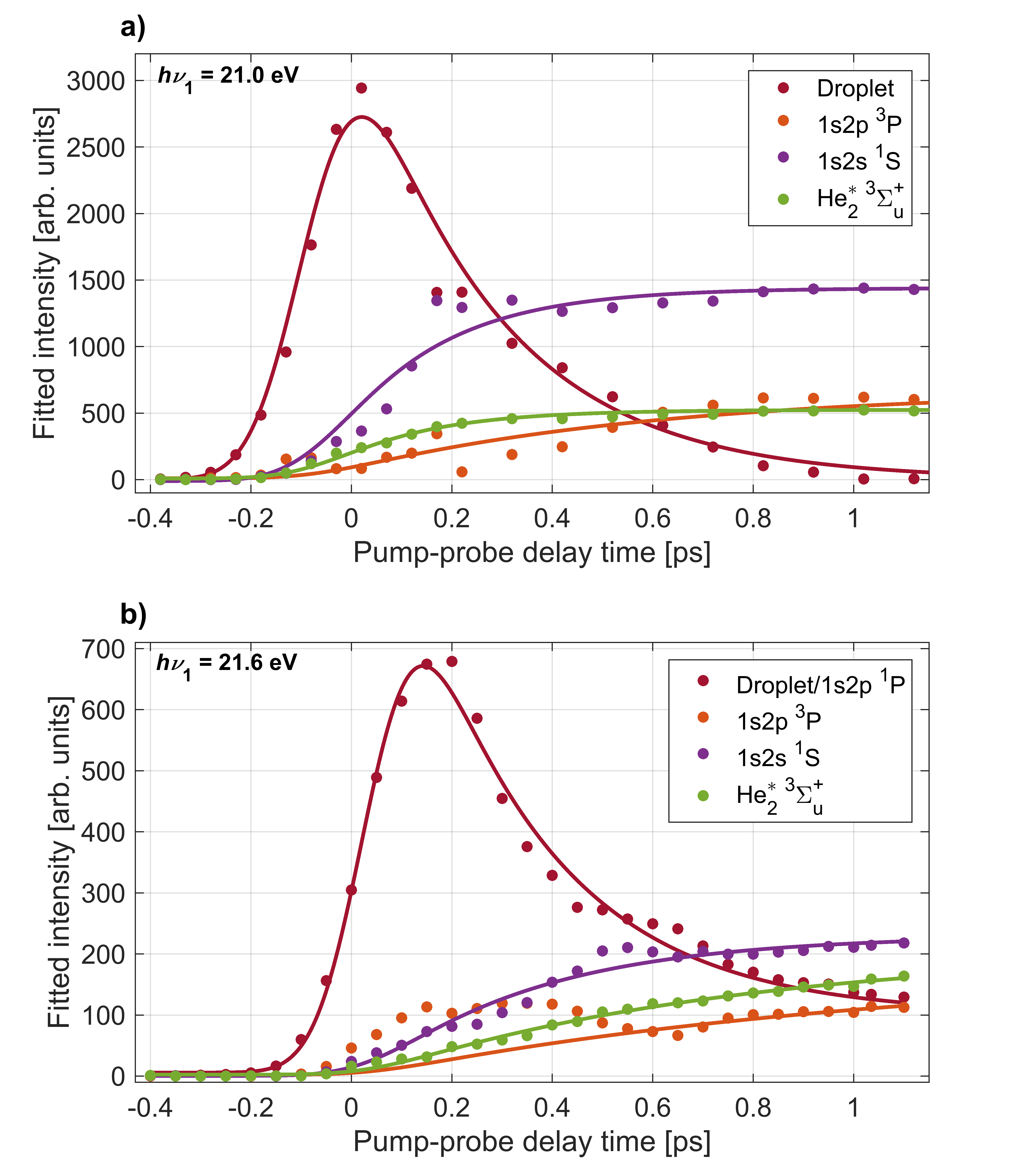}
    \caption{Results of peak fits to the delay-dependent electron spectra recorded at pump photon energies $h\nu_1 = 21.0~$eV [panel a)] and $h\nu_2 = 21.6~$eV [panel b)]. Shown are the integrated areas of Gaussian peaks fitted to the individual components in the electron spectra, labeled in Fig.~\ref{fig:21d6eV+21d0eV}. The largest `Droplet' peak corresponds to the bright feature dominating the electron spectra around zero delay. The smooth lines depict fits of the data with Eqs.~(\ref{eq:Model1}) and (~\ref{eq:Model2}) from which characteristic time constants for the appearance of the individual components are inferred.}
    \label{fig:fit}
\end{figure}

\begin{table}[h]
\begin{tabular}{l|ccc}
Transition & $h\nu_1$ [eV] & $\tau$ [ps] & \begin{tabular}[c]{@{}c@{}}$\tau_{\text{ref}}$ [ps]\\  ($h\nu_1 = 23.8$~eV)\end{tabular} \\ \hline
Decay of $^1S^{(D)}$ & 21.0 & \multicolumn{1}{l}{$0.28 \pm 0.02$} & \multicolumn{1}{l}{} \\
$^1S^{(D)} \rightarrow \,^3P^{(A)}$ & 21.0 & $0.64 \pm 0.17$ &  \\
$^1S^{(D)} \rightarrow \, ^1S^{(A)}$ & 21.0 & $0.21 \pm 0.02$ & $0.78 \pm 0.16$~\cite{asmussen2021unravelling} \\
$^1S^{(D)} \rightarrow \, ^3\Sigma_u^+$ & 21.0 & $0.19 \pm 0.01$ &  \\
Decay of $^1P^{(D)/(A)}$ & 21.6 & \multicolumn{1}{l}{$0.27 \pm 0.03$} & \multicolumn{1}{l}{} \\
$^1P^{(D)} \rightarrow \, ^3P^{(A)}$ & 21.6 & $1.07 \pm 0.45$ &  \\
$^1P^{(D)} \rightarrow \, ^1S^{(A)}$ & 21.6 & $0.34 \pm 0.03$ & \begin{tabular}[c]{@{}c@{}}$0.59 \pm 0.06$~\cite{asmussen2021unravelling}, \\ $\sim0.45$~\cite{Ziemkiewicz2014}\end{tabular} \\
$^1P^{(D)} \rightarrow \, ^3\Sigma_u^+$ & 21.6 & $0.73 \pm 0.06$ & 
\end{tabular}
\caption{Time constants obtained by fitting peaks in the delay-dependent experimental electron spectra with a multi-Gaussian model function (third column). The first column labels the individual components in the electron spectra measured at $h\nu_1=21.0$~eV and $21.6$~eV (second column). The reference values in the fourth column are from the literature.}
\label{fitvalues}
\end{table}

To quantify the time evolution of the spectral features shown in Fig.~\ref{fig:21d6eV+21d0eV}, we fitted the delay-dependent photoelectron spectra with a multi-Gaussian function. The smooth thin lines show the best fits in Figs.~\ref{fig:21d6eV+21d0eV}~b) and d). The resulting peak areas of each Gaussian component are depicted in Figs.~\ref{fig:fit}~a) and b) for the two photon energies $h\nu_1=21.0$ and $21.6$~eV, respectively. In total, the fit function at $21.0$~eV is the sum of 6 individual Gaussian functions to account for 3 sharp atomic lines ($1s2s\,^1S$, $1s2p\,^3P$, $1s2p\,^1P$), two He$_2^*$ molecular lines ($v=0$, $v=1$), and one broadened droplet spectral feature present in the photoelectron spectra shown in Fig.~\ref{fig:21d6eV+21d0eV}~b). At $h\nu_1 = 21.6$~eV (Fig.~\ref{fig:21d6eV+21d0eV}~d)), one additional peak accounting for near-zero kinetic energy electrons generated by autoionization is added to the fit. For the Gaussian functions representing the atomic components, the position fit parameters are constrained to narrow intervals around their well-known excitation energies. Since the photon energy of the probe pulse is insufficient for photoionizing the lowest excited atomic state of He, $1s2s\,^3S$, by one-photon absorption, this state remains undetected in the present study.

The resulting peak areas are plotted in Fig.~\ref{fig:fit} and fitted with the following simple model functions. For the droplet feature, we assume the model function
\begin{equation}
f(t)=\Theta(t-t_0)\left[A + B\exp\left( -(t-t_0)/\tau\right)\right],
\label{eq:droplet}
\end{equation}
where $A$ and $B$ are adjustable constants, $\Theta$ is the Heaviside step function, and $\tau$ is the decay time constant of the signal at positive delays, $t>t_0$. For all other peaks, the model is
\begin{equation}
f(t)=\Theta(t-t_0)C\left[ 1 - \exp\left( -(t-t_0)/\tau\right)\right].
\label{eq:atomic}
\end{equation}
These functions are convoluted with a Gaussian cross-correlation function with a FWHM of 200~fs to account for the finite duration of the pump and probe pulses. This convolution can be carried out analytically and the resulting formulas are given explicitly in the appendix (Eqs.~(\ref{eq:Model1}) and (\ref{eq:Model2})). The best fits of the resulting fit functions to the experimental data are shown in Fig.~\ref{fig:21d6eV+21d0eV} b) and d) as smooth gray lines. At $h\nu_1=21.0$~eV, the $1s2p\,^1P$ atomic state signal is small for all delays and is therefore omitted in the figure. At $h\nu_1=21.6$~eV (Fig.~\ref{fig:21d6eV+21d0eV} b)), the $1s2p\,^1P$ atomic state coincides with the lower edge of the broad droplet feature and therefore is treated as a combined effect from both processes. The local maximum of the $1s2p\,^3P$ peak area around 0.3~ps delay is due to a crosstalk from the dominant $1s2p\,^1P$ droplet signal to the $1s2p\,^3P$ signal at delays $<0.6$~ps. Therefore these data points are excluded from the fit of the  $1s2p\,^3P$ signal.

From these fits we infer the characteristic time constants for each component which are summarized in Table~1, where droplet and atomic states are indicated with the superscripts (D) and (A), respectively. Overall, we observe both similarities and clear differences in the time evolution of the individual components in the electron spectra at the two photon energies $h\nu_1=21.0$~eV and $21.6$~eV. First, the excited droplet state is populated within the range of overlapping pump and probe pulses and subsequently decays within a few hundred fs. At $h\nu_1=21.0$~eV, the droplet excitation, which correlates to the $1s2s\,^1S$ state, reaches its maximum nearly at the zero-point of the delay ($t_\mathrm{max}\approx 20$~fs). Subsequently, it rapidly decays to zero ($\tau\approx 280~$fs) and evolves mostly into a well-separated narrow atomic state of the same symmetry ($1s2s\,^1S$), \textit{cf.} Fig.~\ref{fig:21d6eV+21d0eV}~a), which has a corresponding rise time $\tau\approx 210~$fs. These dynamics can be compared to time-dependent density functional (TDDFT) simulations of the dynamics of a $1s2s\,^1S$ excited He$^*$ atom inside superfluid He~\cite{asmussen2021unravelling}. In these simulations, a void bubble forms around the He$^*$ whose radius was found to increase with an exponential time constant of 0.25~ps, in good agreement with the experimental values measured here. In our previous experiment where He droplets were resonantly excited into the higher-lying droplet absorption band at $h\nu_1=23.7$~eV, relaxation out of the $1s2s\,^1S$ droplet excitation into the $1s2s\,^1S$ atomic state was observed as a secondary process in a step-wise relaxation starting from a broad excitation correlating to the He $1s3p$ and $1s4p$ droplet states~\cite{asmussen2021unravelling}. There, a longer relaxation time $\approx 780$~fs was measured. We ascribe the slower relaxation measured in those experiments to a weaker coupling of the $1s2s\,^1S$ excitation to the surrounding He$^*$ when this state is populated indirectly by relaxation out of a higher level. In other words, as a high-lying state is excited, a bubble opens up around the He$^*$ over the course of electronic relaxation, thereby reducing the orbital overlap of the He$^*$ with surrounding He atoms. 

At $h\nu_1=21.6$~eV, where the He $1s2p\,^1P$ droplet state is excited, the maximum intensity of the corresponding photoelectron signal is reached at a later delay time $t_\mathrm{max}=150$~fs before dropping with a similar decay time $\tau\approx 270~$fs as compared to the dynamics at $h\nu_1=21.0$~eV. Subsequently, the droplet state evolves continuously into the atomic $1s2p\,^1P$ state which appears as a narrow band at the lower edge of the droplet feature in Fig.~\ref{fig:21d6eV+21d0eV}~c). The most abundant atomic state populated by relaxation is $1s2s\,^1S$ with a corresponding rise time $\tau\approx 340~$fs. Relaxation into this state requires a decrease of the electron energy by up to 1~eV and a change of symmetry of the electronic state from $1s2p\,^1P$ to $1s2s\,^1S$. This explains the wide gap between these two bright features in the time-resolved electron spectra, see Fig.~\ref{fig:21d6eV+21d0eV}~c), and the slower relaxation dynamics as compared to the direct excitation of the $1s2s\,^1S$ droplet state at $h\nu_1=21.0$~eV [Fig.~\ref{fig:21d6eV+21d0eV}~a)]. In previous experiments where the higher $1s3p/1s4p$ droplet band was excited, relaxation from the $1s2p\,^1P$ droplet excitation to the $1s2s\,^1S$ atomic state was measured to be slower, $\tau\approx 590~$fs~\cite{Ziemkiewicz2014}. In another experiment using XUV pulses generated by HHG at nearly the same photon energy, $\tau\approx 450$~fs was measured~\cite{Ziemkiewicz2014}. This confirms the trend that secondary steps within a relaxation cascade are slower, as the coupling of the excited state to the droplet weakens over the course of electronic relaxation and simultaneous bubble formation.


The emerging atomic-like triplet states $1s2p\,^3P$ and the excimer state $^3\Sigma_u^+$ have considerably longer rise times, see Table~1. Only the $^3\Sigma_u^+$ state at $h\nu_1 = 21.0$~eV appears to be populated at the same rate as the initial relaxation of the $1s2s\,^1S$ state. This may be due to its strong overlap with the $1s2s\,^1S$ state in the electron spectra which introduces a large uncertainty in the peak fitting procedure. The large uncertainty of the $1s2p\,^3P$ time constant at both photon energies is due to the spectral overlap of this component with the low-energy tail of the broad droplet feature at short time delays. The slower appearance time of triplet states is likely related to the fact that a spin flip is needed since optically excited states have singlet symmetry. Previous fluorescence measurements have suggested that triplet states are only formed by electron-ion recombination when exciting into the autoionizing states at $h\nu>23$~eV~\cite{VonHaeften1997}. However, indications for the population of triplet states by relaxation even out of the lower-lying $1s2s$ and $1s2p$ states have been found; Penning ionization of alkali metal atoms attached to He nanodroplets occurred from both the He $1s2s\,^1S$ state and, to a small extent, the $1s2s\,^3S$ state, after excitation into the lower absorption band of the droplet at $h\nu=21.6$~eV~\cite{BenLtaief2019}. Spin quenching was also observed for barium atoms attached to the surface of argon clusters~\cite{awali2016multipronged}. Thus, we conclude that He droplets are capable of inducing spin quenching of excited states to some degree as well, and the time scale is 600-1100~fs.

Additionally, we have performed similar measurements at $h\nu_1 =24.5$~eV which is near the He atomic ionization threshold. The results are shown in Fig.~\ref{fig:24d6eV}. The most prominent feature in the electron spectra at this photon energy for all delays is a distribution of low-energy electrons which falls off exponentially in the range $eKE=0$-2~eV (not shown). This electron distribution is generated by the XUV pump pulse and is due to the autoionization of He droplets excited above their adiabatic ionization energy~\cite{Peterka:2003,Peterka:2007}. The low-energy part of this distribution appears as a negative signal in Fig.~\ref{fig:24d6eV} since the pump-only spectrum is subtracted from all shown photoelectron spectra and the probe pulse efficiently depletes this channel by photoionizing the excited-state electrons into the continuum. 

On top of this large signal, we detect photoelectrons emitted by 1+1' photoionization with energies $eKE=4.5$~eV. As it is a weak positive signal on a large background, the signal-to-noise ratio is low and experimental artifacts are seen in the spectra as horizontal stripes. The energy resolution may have been somewhat compromised having used an additional retarding voltage near the interaction region in the attempt to reduce the contribution of low-energy electrons in the spectra. Nevertheless, a few trends are clearly discernible. Similar to the time-dependent spectra recorded at $h\nu_1=21.0$~eV and $21.6$~eV discussed above, as well as the ones previously recorded at $h\nu_1=23.7$~eV~\cite{asmussen2021unravelling}, the initially excited droplet state (bright red distribution around $t=0$, $eKE=4.5~$eV) relaxes into lower-lying states, in particular the $1s2s\,^1S$ band around $eKE\approx 0.7~$eV, within $\sim 150~$fs. However, in this case, a substantial amount of population remains in higher-lying states at $eKE>3~$eV. As these excited states continue to undergo autoionization, depletion of these states by the probe pulse quenches the autoionization process in the entire shown range of delays (negative signal near $eKE=0$). In our previous measurement at $h\nu_1=23.7$~eV~\cite{asmussen2021unravelling}, depletion of the autoionization signal was only visible in the range of overlapping pump and probe pulses around $t=0$, in line with the observation of fast relaxation of electron energies to states with $eKE<3~$eV.
Such a fast reappearance time of the autoionization signal following depletion of excitation at $h\nu_1=23.7$~eV was also seen in experiments using XUV pulses generated by HHG~\cite{Kornilov2010,Ziemkiewicz2015}.
Thus, we conclude that droplet excitation at $h\nu_1>23~$eV (correlating to He atomic states with principal quantum numbers $n\geq 3$~\cite{Kornilov2011}) relax more slowly if populated by prior relaxation from even higher excited states. It is likely that bubble formation initiated by the electronic excitation reduces the coupling of the excited state to the droplet over the course of relaxation, thereby gradually slowing down any subsequent electronic relaxation steps.

\section{Conclusions}
In summary, we have performed time-resolved, high-resolution photoelectron spectroscopy of resonantly excited He nanodroplets. In the XUV-pump and UV-probe electron spectra, we observed various distinct features due to two-color photoionization, ATI, and autoionization of multiply excited droplets by ICD. In particular, thanks to the improved resolution and a higher signal-to-noise ratio compared to previous measurements of electron spectra, we could observe new, ultrafast relaxation channels of the excited droplets into triplet atomic states, indicating efficient droplet-induced spin-relaxation, as well as the formation of the first excimer state of He$_2^*$. The convergence of these features to atomic-level energies on a timescale of a few 100~fs indicates ultrafast localization of the excitation at quasi-free He$^*$ atoms and He$_2^*$ excimers residing in void bubbles or ejected from the droplets. From multi-Gaussian peak fits of the delay-dependent spectra, we inferred the relaxation-time constants for the individual final states. Significantly different relaxation dynamics were seen for excitation of the two lowest optically accessible He droplet states $1s2s\,^1S$ and $1s2p\,^1P$, despite them overlapping in the static absorption spectrum~\cite{Joppien1993}. In both cases, we find that excited He droplets partially relax into previously unobserved triplet states, indicating efficient He droplet-induced spin relaxation. 

Thus, He nanodroplets were found to turn into highly reactive and dissipative systems upon resonant excitation, featuring complex ultrafast relaxation dynamics including electronic-state hopping and spin flipping, the formation of bubbles and excimers, and few-body autoionization. Future time-resolved high-resolution photoelectron spectroscopic measurements of other types of pure or doped noble-gas clusters and nanodroplets will shed more light onto the peculiar properties of superfluid He nanodroplets on the one hand, and on more general aspects of the relaxation dynamics of nanoparticles irradiated by resonant UV or XUV radiation on the other hand.

\section{Author contribution}
A.C.L., B.B., M.B., C.C., M.D.F., S.H., S.R.K., A.N., N.P., O.P., P.R., N.B., and M. M. performed the experiments. J.D.A., S.D., L.G., M.F.K., D.M., S.M., K.C.P., E.S., and F.S. aided remotely in the analysis and interpretation of the experimental results. J.D.A. analyzed the data. A.C.L., J.D.A., and M.M. wrote the manuscript with input from all the co-authors.

\section{Conflict of interest}
The authors declare no conflicting interests.

\section{Acknowledgements}
We gratefully acknowledge the support of the Chemical Sciences, Geosciences and Biosciences Division, Office of Basic Energy Sciences, Office of Science, U.S. Department of Energy, Grant No. DESC0012376 and DESC0063, the German Science Foundation (DFG) through project STI 125/19-2, and the Danish Agency for Science, Technology, and Innovation through the instrument center DanScatt. A.C.L. gratefully acknowledges the support of Charles and Peggy Hardin. M.F.K. is grateful for support by the Max Planck Society via the Max Planck Fellow program. S.R.K. gratefully acknowledges the financial support from the Min. of Education, Govt of India through the scheme for promotion of research and academic collaboration and the Institute of Eminence programmes, from the Core Research Grant, Indo-Elettra and India-DESY schemes of the Dept. Of Science and Technology, Govt. Of India and from the Max Planck Society, Germany. M.M. is grateful for financial support from the Carlsberg Foundation.

\section{Appendix}
The fit functions used for modelling the experimental data shown in Fig.~\ref{fitvalues} are given by the convolution of the model functions Eq. (\ref{eq:droplet}) and Eq. (\ref{eq:atomic}) with a Gaussian function that accounts for the cross-correlation of the pump and probe pulses,
\[
M(t)=\frac{1}{\sqrt{2\pi}\sigma}e^{-\frac{(t-t_0)^2}{2\sigma^2}}.
\]
The standard deviation $\sigma$ is related to the FWHM, which is 200~fs, by $\sigma = \mathrm{FWHM}/\sqrt{8\ln{2}}$.
For model function~(\ref{eq:droplet}) the convolution can be written out as 
\begin{eqnarray}
I(t)&=&\int_{-\infty}^\infty f(t') M(t'-t+t_0)dt' \label{eq:AppConv} \nonumber \\ 
     &=& \frac{A}{2}\left[ 1+\mathrm{erf}\left(\frac{t-t_0}{\sqrt{2}\sigma}\right)\right] \nonumber \\ 
     &+& \frac{B}{2}\exp\left(\frac{\sigma^2 -2\tau (t-t_0)}{2\tau^2}
     \right)\mathrm{erfc}\left(\frac{\sigma^2 - \tau (t-t_0)}{\sqrt{2}\sigma\tau}\right). \label{eq:Model1}
\end{eqnarray}

For model function~(\ref{eq:atomic}) it is 
\begin{eqnarray}
I(t) &=& \frac{C}{2}\left[ 1 + \mathrm{erf}\left(\frac{t-t_0}{\sqrt{2}\sigma}\right)\right] \nonumber\\ 
&-& \frac{C}{2}\exp\left(\frac{\sigma^2 -2\tau (t-t_0)}{2\tau^2}\right)\mathrm{erfc}\left(\frac{\sigma^2 - \tau (t-t_0)}{\sqrt{2}\sigma\tau}\right). \label{eq:Model2}
\end{eqnarray}
Here, $\mathrm{erf}(z)=2\int_0^z\exp\left(-t^2\right)dt/\sqrt{\pi}$ denotes the error function and $\mathrm{erfc}(z)=1-\mathrm{erf}(z)$.
\balance

\bibliography{Relaxation_Bib}

\bibliographystyle{rsc}

\end{document}